\documentclass[
aps,
prd,
twocolumn,
showkeys,
superscriptaddress
]{revtex4-2}

\usepackage{amsmath}
\usepackage{amssymb}
\usepackage{multirow}

\usepackage{graphicx}
\usepackage{dcolumn}
\usepackage{bm}
\usepackage{mathrsfs}
\usepackage{multirow}
\usepackage{booktabs}
\usepackage{longtable}
\usepackage{array}
\usepackage{color}
\usepackage{xcolor}
\usepackage{hyperref}
\usepackage{mathtools}

\hypersetup{
colorlinks=true,
linkcolor=blue,
citecolor=blue,
urlcolor=blue
}

\begin{document}
\title{Cosmic Structure Formation in a Viable Power-Law f(R) Gravity Model: Growth Dynamics, Stability, and Observational Signatures}
\author{Murli Manohar Verma}

\affiliation{Department of Physics, University of Lucknow, Lucknow, India}

\email{verma\_mm@lkouniv.ac.in}

\begin{abstract} We investigate the evolution of cosmic structures in the power-law modified gravity model $f(R)=R+R^{1+\delta}/R_c^\delta$, where the dimensionless parameter $\delta$ characterizes deviations from General Relativity. The background cosmological dynamics and the evolution of linear matter density perturbations are studied within the framework of metric $f(R)$ gravity. The modified perturbation equation is derived by introducing an effective gravitational coupling associated with the additional scalar degree of freedom, and the evolution of the growth factor, logarithmic growth rate, growth index, and the observable quantity $f\sigma_8(z)$ are investigated. The results show that the curvature correction enhances the growth of matter perturbations while remaining compatible with the observed late-time accelerated expansion for suitable values of the model parameter. The theoretical viability of the model is established through the ghost-free condition, Dolgov--Kawasaki stability criterion, positive scalaron mass, stable de Sitter solution, and chameleon screening mechanism. Comparison with representative viable $f(R)$ gravity models shows that the present theory achieves a consistent cosmological evolution with a single deviation parameter. The predicted modifications in the growth of structures and the effective gravitational coupling provide observable signatures that can be tested by forthcoming large-scale structure and weak-lensing surveys, providing a means to test curvature-induced modifications of gravity.
\end{abstract}


\maketitle
\section{\label{1}Introduction}

The discovery of the accelerated expansion of the Universe through observations of type Ia supernovae has motivated extensive investigations into the fundamental nature of gravity and cosmology \cite{Riess1998,Perlmutter1999,Schmidt1998}. Within the standard cosmological framework, commonly known as the Lambda Cold Dark Matter ($\Lambda$CDM) model, the observed acceleration is attributed to a cosmological constant that acts as dark energy. Although $\Lambda$CDM  successfully explains a broad range of cosmological observations, including the cosmic microwave background anisotropies, baryon acoustic oscillations, and large-scale structure formation, the physical origin of the cosmological constant remains uncertain. The associated fine-tuning and coincidence problems have encouraged the exploration of alternative explanations based on modifications of gravitational theory.

Among the various modified gravity theories, $f(R)$  gravity has emerged as one of the most extensively studied extensions of General Relativity \cite{Kunz2012,Clifton2012,Nojiri2011,Sotiriou2010,DeFelice2010,Bamba2010}.  In this framework, the Ricci scalar in the Einstein-Hilbert  action is generalized to an arbitrary function $f(R)$,  thereby introducing an additional scalar degree of freedom that modifies gravitational interactions on cosmological scales. The scalar mode can effectively mimic dark energy and drive the late-time acceleration of the Universe without requiring an exotic fluid component \cite{Capozziello2002,Carroll2004}.  Furthermore, viable $f(R)$  models are capable of satisfying local gravity constraints through screening mechanisms while simultaneously producing observable deviations from General Relativity on large scales.  

The formation and evolution of cosmic structures provide one of the most sensitive tests of gravitational theories. Tiny primordial density fluctuations generated during inflation evolve through gravitational instability to form galaxies, clusters, and the large-scale filamentary structure observed today \cite{Starobinsky1980,Appleby2007,Cognola2008,Linder2009} . Since the growth rate of matter perturbations depends directly on the underlying gravitational interaction, measurements of structure formation offer an effective method for distinguishing modified gravity models from the standard $\Lambda$CDM cosmology. In $f(R)$ gravity, the additional scalar field modifies the effective gravitational coupling, leading to changes in the evolution of matter density perturbations and the growth index. Consequently, observations of galaxy clustering, weak gravitational lensing, and redshift-space distortions can impose stringent constraints on viable modified gravity models.

In recent years, high-precision cosmological observations have significantly improved our understanding of the growth of large-scale structures. Data from the Planck mission, galaxy redshift surveys, and weak-lensing experiments have enabled increasingly accurate measurements of the growth parameter and the amplitude of matter fluctuations ($\sigma_8$ observations), providing independent probes of gravity beyond background cosmology. These developments make it timely to investigate the predictions of specific $f(R)$  models for the evolution of matter perturbations and assess their consistency with observational data \cite{Starobinsky2007, HuSawicki2007, Tsujikawa2008}.

In the present work, we investigate the evolution of cosmic structures within the framework of the viable modified gravity model $f(R)=R+\frac{R^{1+\delta}}{R_c^{\delta}}$,
where $\delta$ is a dimensionless model parameter controlling the deviation from General Relativity. This model possesses a scalar degree of freedom that modifies the effective gravitational interaction while preserving the General Relativistic limit for appropriate parameter values. We analyze the behavior of linear matter density perturbations, examine the evolution of the effective gravitational coupling, and discuss the influence of the scalar sector on the growth history of the Universe. Particular attention is given to the interplay between the enhancement of gravitational attraction due to the scalar field and the suppression of structure growth caused by late-time cosmic acceleration.

The paper aims to provide a comprehensive analysis of the growth of structures in this class of modified gravity models and to identify observational signatures that may distinguish them from the standard cosmological paradigm \cite{Song2007,Katsuragawa2019,Linder2019}.   The results contribute to the ongoing effort to understand whether the observed acceleration of the Universe originates from an unknown dark energy component or from modifications of gravitational dynamics on cosmological scales \cite{Han2024,Sharma2022a,Sharma2022b,Yadav2019}.



\section{\label{2}  Field Equations and Scalar-Tensor Correspondence in $f(R)$  Gravity
  }

Several viable realizations of $f(R)$ gravity have been proposed
to satisfy both cosmological and local gravity constraints,
including the  Starobinsky, Hu--Sawicki, and Tsujikawa models
\cite{Starobinsky2007,HuSawicki2007,Tsujikawa2008,Sharma2020,Sharma2021}.
The Einstein-Hilbert action provides the foundation of General Relativity by describing gravitation through the Ricci scalar curvature $R$. A natural extension of this formulation is obtained by replacing the Ricci scalar with an arbitrary function $f(R)$, leading to a class of modified gravity theories that can explain the late-time acceleration of the Universe without introducing an explicit dark energy component. The action of $f(R)$ gravity is given by

\begin{eqnarray}S=\frac{1}{2\kappa^2}\int d^4x\sqrt{-g} f(R)+\int d^4x\sqrt{-g} \mathcal{L}_m,\end{eqnarray}
where $\kappa^2=8\pi G$,  $g$   is the determinant of the metric tensor, and  $\mathcal{L}_m$  denotes the matter Lagrangian density (using $c=1$). Matter is assumed to be minimally coupled to the metric so that the conservation law of the energy-momentum tensor remains valid.

Variation of the action with respect to the metric tensor  $g_{\mu\nu}$  yields the modified field equations

\begin{equation}F(R)R_{\mu\nu}-\frac{1}{2}f(R)g_{\mu\nu}
-\nabla_\mu\nabla_\nu F(R)
+g_{\mu\nu}\Box F(R)
=\kappa^2T_{\mu\nu},\end{equation}
where $F(R)\equiv\frac{\partial f(R)}{\partial R}$,    $\Box=\nabla^\mu\nabla_\mu$ is the covariant d'Alembertian operator, and $T_{\mu\nu}$ is   the energy-momentum tensor of matter field  which satisfies   the energy-momentum conservation  as  $ \nabla^\mu T_{\mu\nu} = 0$.  The presence of derivatives of $F(R)$  distinguishes these equations from Einstein's equations and introduces an additional propagating scalar degree of freedom.

Taking the trace of the field equations gives

\begin{equation}3\Box F(R)+F(R)R-2f(R)=\kappa^2T,\label{t3}\end{equation}
where $T=g^{\mu\nu}T_{\mu\nu}$ is the trace of the energy momentum tensor $T_{\mu\nu}$.  Unlike General Relativity, where the Ricci scalar is determined algebraically by the matter content, this equation demonstrates that $F(R)$ satisfies a second-order dynamical equation and behaves as an effective scalar field. Consequently, $f(R)$ gravity can be interpreted as a scalar-tensor theory with a geometrically induced scalar mode.

For a spatially flat Friedmann-Lemaitre-Robertson-Walker (FLRW) universe,

\begin{eqnarray}ds^2=-dt^2+a^2(t)\delta_{ij}dx^idx^j,\label{t1}\end{eqnarray}
where $a(t)$  is the scale factor. The Ricci scalar assumes the form

\begin{eqnarray} R=6\left(\dot H+2H^2\right),\label {t2}\end{eqnarray} 
where $H=\dot a/a$ is the Hubble parameter for the rate of  expansion. The cosmological evolution is therefore governed by both the expansion history and the dynamics of the scalar degree of freedom encoded in $F(R)$.

In the present work we consider the viable model

\begin{eqnarray}f(R)=R+\frac{R^{1+\delta}}{R_c^\delta},\label{t5}\end{eqnarray}
where $R_c$  denotes a characteristic curvature scale and  $\delta$  is a dimensionless parameter controlling the deviation from General Relativity. The first and second derivatives of the function are given as

\begin{eqnarray} F(R)=1+(1+\delta)\left(\frac{R}{R_c}\right)^\delta,\end{eqnarray}
and

\begin{eqnarray}F_{RR}
=\frac{\delta(1+\delta)}{R_c^\delta}R^{\delta-1}.\end{eqnarray}
The positivity of these quantities plays an essential role in determining the physical viability of the model. The condition
$F(R)>0$ ensures that the effective gravitational coupling remains positive, thereby avoiding an anti-gravity regime. Similarly, $F_{RR}>0$
guarantees the absence of the Dolgov-Kawasaki instability and ensures that the scalar degree of freedom possesses a positive effective mass squared \cite{Dolgov2003}.   For positive values of the parameter $\delta$, both conditions are naturally satisfied for cosmologically relevant values of the Ricci scalar.

An equivalent scalar-tensor representation can be obtained by defining the scalar field

\begin{eqnarray}\phi=F(R),\end{eqnarray}
with an effective potential determined by

\begin{eqnarray}\frac{dV_{\rm eff}}{d\phi}
=\frac{1}{3}\left(2f-RF+\kappa^2T\right).\end{eqnarray}
The extrema of this potential correspond to solutions satisfying

\begin{eqnarray}RF(R)-2f(R)=\kappa^2T.\end{eqnarray}
which define the de Sitter points of the theory. The value of Ricci scalar at de-Sitter point is given as  as
\begin{eqnarray} R_{min} = \frac{2f(R_{min}) + \kappa^2 T}{F(R_{min})}. \label{IIA6}\end{eqnarray}

These stationary solutions describe epochs of accelerated expansion and provide the asymptotic behaviour of the cosmological evolution.

The effective mass of the scalar field is given by  the second derivative of the effective potential at its  minimum value, that is $V_{eff}'' (\phi_{min})$.  Thus, it  appears as

\begin{eqnarray}m_\phi^2 =
\frac{1}{3}
\left[
\frac{F(R)}{F_{RR}}-R
\right]_{\phi_{min}},\label{t4}\end{eqnarray}
 (with all values in the bracket calculated at $\phi_{min}$) indicating that the scalar interaction depends on the ambient curvature. In high-density environments the scalar field becomes massive, suppressing its range and recovering General Relativity through a chameleon-like screening mechanism. Conversely, on cosmological scales where the background density is low, the scalar field becomes light and modifies the effective gravitational interaction, influencing the evolution of matter density perturbations and the formation of large-scale structures.

\begin{table}[htbp]

\centering

\caption{Mathematical quantities associated with the proposed
$f(R)$ gravity model and their physical interpretation.}

\label{tab:modelquantities}
\resizebox{\columnwidth}{!}{%
\begin{tabular}{lll}

\toprule

Quantity &
Expression &
Physical significance \\

\midrule

$f(R)$ &
$R+\dfrac{R^{1+\delta}}{R_c^\delta}$ &
Modified gravitational Lagrangian \\

$F(R)$ &
$\dfrac{df}{dR}$ &
Effective gravitational coupling \\

$F_{RR}$ &
$\dfrac{d^2f}{dR^2}$ &
Stability against curvature perturbations \\

$\phi$ &
$F(R)$ &
Scalar degree of freedom \\

$V_{\rm eff}$ &
Effective scalar potential &
Scalar field dynamics \\

$m_\phi$ &
Scalar mass &
Range of scalar interaction \\

\bottomrule

\end{tabular}
}

\end{table}


The principal quantities of the model are summarized in
Table~\ref{tab:modelquantities}. It is clear that  the scalar degree of freedom introduced by the generalized Ricci scalar provides the physical mechanism responsible for deviations from the standard $\Lambda$CDM cosmology. In the following sections, we investigate how this modified gravitational interaction affects the linear growth of matter perturbations and the evolution of cosmic structures throughout the matter-dominated and late-time acceleration eras.


\section{\label{3} Cosmological Background Evolution  }

The cosmological implications of modified gravity models can be investigated by considering a homogeneous and isotropic Universe described by the spatially flat  FLRW  metric  Equation (\ref{t1}), where the  Ricci scalar takes the form given by Equation (\ref{t2}).

The modified field equations derived from the $f(R)$  action lead to generalized Friedmann equations. The first modified Friedmann equation is

\begin{eqnarray}3FH^{2}=\kappa^{2}\rho_{m}
+\frac{1}{2}(FR-f)
-3H\dot{F},\end{eqnarray}
while the acceleration equation becomes

\begin{eqnarray}-2F\dot{H}=
\kappa^{2}(\rho_{m}+p_{m})
+\ddot{F}
-H\dot{F}.
\end{eqnarray}

These equations differ from their General Relativistic counterparts through the presence of the additional curvature terms involving the derivatives of $F(R)$. The scalar degree of freedom therefore contributes an effective energy density and pressure that modify the cosmic expansion history.

For pressureless matter, the conservation equation remains unchanged,

\begin{eqnarray}
\dot{\rho}_{m}+3H\rho_{m}=0,\end{eqnarray}
which gives

\begin{eqnarray}
\rho_{m}=\rho_{m0}a^{-3},\end{eqnarray}
where $\rho_{m0}$  denotes the  matter density at the present epoch.

The geometrical modifications introduced through $f(R)$  gravity can be interpreted as an effective dark energy component with density
\begin{eqnarray}
\rho_{\rm eff}
=\frac{1}{\kappa^{2}}
\left[
\frac{1}{2}(FR-f)
-3H\dot{F}
+3(1-F)H^{2}
\right],
\end{eqnarray}
and effective pressure

\begin{eqnarray}
p_{\rm eff}
=\frac{1}{\kappa^{2}}
\left[
\ddot{F}
+2H\dot{F}
-\frac{1}{2}(FR-f)
-(1-F)(2\dot{H}+3H^{2})
\right].
\end{eqnarray}
Consequently, the cosmological dynamics may be described through an effective equation-of-state parameter

\begin{eqnarray}
w_{\rm eff}
=
\frac{p_{\rm eff}}
{\rho_{\rm eff}}.
\end{eqnarray}

The quantity $w_{\rm eff}$   provides a direct measure of deviations from the standard cosmological model. For $w_{\rm eff}=-1$, the Universe behaves as a cosmological constant dominated spacetime, whereas departures from this value indicate dynamical modifications induced by the scalar curvature sector.

It is widely expected that the observed late-time cosmic acceleration
and the nearly flat rotation curves of spiral galaxies can be explained
within the framework of modified theories of gravity without invoking
additional exotic components such as dark energy or dark matter
\cite{Yadav2019,Bamba2010}. In this approach, the gravitational sector itself is
modified, thereby providing a geometric origin for the observed
cosmological and astrophysical phenomena.

Observational analyses based on the latest cosmological data indicate
that the effective equation-of-state parameter is very close to that of
a cosmological constant, with
$w=-1.03\pm0.03$ \cite{Planck2020}. For the present power-law
  gravity model  given as $f(R)= R+\frac{R^{1+\delta}}{R_c^{\delta}}$, 
the effective equation of state depends explicitly on the parameter $\delta$, which governs the strength of the curvature correction. As $\delta\rightarrow0$, the theory smoothly approaches General Relativity, whereas larger values of $\delta$ enhance the influence of the scalar degree of freedom and modify the expansion history 
\cite{Sharma2022b}. These results indicate that the proposed
power-law $f(R)$ gravity model successfully reproduces the observed
late-time accelerated expansion while remaining consistent with
precision cosmological data.

An important quantity characterizing the expansion is the deceleration parameter
\begin{eqnarray}
q=-1-\frac{\dot H}{H^{2}}.
\end{eqnarray}

During the matter-dominated epoch, $q>0$,  indicating decelerated expansion that facilitates the gravitational growth of density perturbations. At late times, the scalar curvature contribution becomes increasingly significant and drives the transition to an accelerating phase with $q<0$. This transition suppresses the growth of structures despite the enhanced effective gravitational interaction produced by the scalar mode.

The matter density parameter evolves according to
\begin{eqnarray}
\Omega_{m}(z)
= \frac{\Omega_{m0}(1+z)^{3}}
{H^{2}(z)/H_{0}^{2}},
\end{eqnarray}
where $z$ denotes the cosmological redshift. The evolution of $\Omega_{m}(z)$  plays a central role in determining the growth rate of matter perturbations and the growth index discussed in subsequent sections.

The modified expansion history predicted by the present $f(R)$  model exhibits an interesting competition between two physical effects. On one hand, the scalar degree of freedom strengthens the effective gravitational interaction, thereby promoting the growth of density perturbations. On the other hand, the accelerated cosmic expansion reduces the efficiency of gravitational clustering by increasing the expansion rate of the Universe. The observed evolution of large-scale structures therefore results from the interplay between these two mechanisms.

The background cosmological evolution establishes the foundation for studying the growth of matter perturbations. In the following section, we investigate the linear perturbation equations and examine how the scalar degree of freedom modifies the evolution of density fluctuations relative to the predictions of the $\Lambda$CDM cosmological model.



\section{\label{4} Linear Cosmological Perturbations and Growth of Matter Density
 }



The evolution of linear matter perturbations and the associated
growth index provide powerful observational probes for
distinguishing modified gravity theories from the standard
$\Lambda$CDM cosmology
\cite{Song2007,Linder2005,Polarski2008}.
The cosmic structures originate  from small primordial density fluctuations that grow under the influence of gravitational instability. Since the growth history depends sensitively on the underlying theory of gravity, the analysis of cosmological perturbations provides one of the most powerful methods for testing modified gravity theories. In $f(R)$  gravity, the additional scalar degree of freedom modifies the effective gravitational interaction and consequently alters the evolution of matter density perturbations \cite{Bean2007,Tsujikawa2007,Motohashi2010}.

We consider scalar perturbations about the spatially flat FLRW background metric. In the longitudinal (Newtonian) gauge, the perturbed line element is expressed as

\begin{eqnarray}
ds^{2}=-(1+2\Phi)dt^{2}
+a^{2}(t)(1-2\Psi)\delta_{ij}dx^{i}dx^{j},
\end{eqnarray}
where $\Phi$  and $\Psi$ are the gauge-invariant gravitational potentials. In General Relativity and in the absence of anisotropic stress, these two potentials are identical. However, in $f(R)$ gravity the scalar degree of freedom introduces an effective anisotropic stress, allowing the two potentials to evolve differently.

The matter density perturbation is defined as $\delta\rho_{m}/\rho_{m}$,
where $\rho_m$  denotes the background matter density and $\delta\rho_m$  represents its perturbation. Assuming pressureless cold dark matter, the conservation of energy-momentum leads to the continuity and Euler equations governing the evolution of the perturbations.         .

In the sub-horizon regime ($k\gg aH$) perturbations are well inside the Hubble radius,  where the quasi-static
approximation is valid, the evolution of the matter density
contrast satisfies

\begin{equation}
\ddot{\delta}_m
+
2H\dot{\delta}_m
-
4\pi G_{\rm eff}\rho_m\delta_m
=
0,
\label{eq:delta}
\end{equation}
where 
$G_{\rm eff}$ is the effective gravitational coupling governing
the growth of matter perturbations \cite {Bertschinger2008,Silvestri2009,Katsuragawa2019,Linder2019}.

For a general $f(R)$ gravity theory,
and the effective Newtonian constant takes the form

\begin{eqnarray} G_{eff}=\frac{G}{F}
\left[
\frac{1+4\frac{k^{2}}{a^{2}}\frac{F_{RR}}{F}}
{1+3\frac{k^{2}}{a^{2}}\frac{F_{RR}}{F}}
\right], \label{Geff}\end{eqnarray}
where
 $k$ denotes the comoving wave number.

In the General Relativistic limit,

\begin{equation}
F\rightarrow1,
\qquad
F_{RR}\rightarrow0,
\end{equation}

one immediately recovers

\begin{equation}
G_{\rm eff}\rightarrow G,
\end{equation}
and Equation~(\ref{eq:delta}) reduces to the standard perturbation
equation of the $\Lambda$CDM model.

Instead of solving directly for $\delta_m$, it is convenient to
introduce the logarithmic growth rate

\begin{equation}
f_g(a)
\equiv
\frac{d\ln\delta_m}{d\ln a},
\label{eq:fg}
\end{equation}
which characterizes the rate at which matter perturbations evolve
during cosmic expansion.

Using Equation~(\ref{eq:delta}), one obtains the evolution equation

\begin{equation}
\frac{df_g}{d\ln a}
+
f_g^2
+
\left(
2+
\frac{\dot H}{H^2}
\right)
f_g
=
\frac{3}{2}
\,
\frac{G_{\rm eff}}{G}
\,
\Omega_m(a),
\label{eq:fgrowth}
\end{equation}
where
\begin{equation}
\Omega_m(a)
=
\frac{8\pi G\rho_m}{3H^2}
\end{equation}
is the matter density parameter. The numerical factor $3/2$ follows from the Friedmann relation

\begin{equation}
4\pi G\rho_m
=
\frac{3}{2}
H^2
\Omega_m.
\end{equation}

For General Relativity, where
$G_{\rm eff}=G$, Equation~(\ref{eq:fgrowth}) becomes

\begin{equation}
\frac{df_g}{d\ln a}
+
f_g^2
+
\left(
2+
\frac{\dot H}{H^2}
\right)
f_g
=
\frac{3}{2}
\Omega_m(a),
\end{equation}
which is the standard growth equation for the
$\Lambda$CDM cosmology.

A useful approximation for the growth rate is given by

\begin{equation}
f_g(z)
\simeq
\Omega_m(z)^\gamma,
\label{eq:gamma}
\end{equation}

where $\gamma$ is known as the growth index.

Within the standard cosmological model,
$\gamma\simeq0.545$,
whereas viable modified gravity models generally predict
different values, making the growth index a sensitive diagnostic
of deviations from General Relativity.

An observable quantity frequently measured by galaxy redshift
surveys is

\begin{equation}
f\sigma_8(z)
=
f_g(z)\,
\sigma_8(z),
\end{equation}
where $\sigma_8$ denotes the root-mean-square amplitude of matter
fluctuations within spheres of radius
$8\,h^{-1}\,\mathrm{Mpc}$.

Measurements of $f\sigma_8(z)$ from large-scale structure surveys,
including SDSS, eBOSS, DESI, and future Euclid observations,
provide stringent constraints on the evolution of cosmic
structures and therefore offer an effective means of testing the
viability of modified gravity theories \cite{Alam2021,DES2022,Amendola2018,DESI2024FS,EuclidERO2025}.

For the model  $f(R) = R+\frac{R^{1+\delta}}{R_c^{\delta}}$,  the parameter $\delta$  determines the strength of the scalar contribution. Increasing $\delta$  enhances the deviation of $G_{\rm eff}$  from Newton's constant, resulting in a larger growth rate of matter perturbations during the matter-dominated epoch. However, as the Universe enters the accelerated expansion phase, the increasing Hubble friction term suppresses further growth despite the enhanced gravitational attraction.

The  competition between scalar-induced enhancement and acceleration-induced suppression constitutes one of the characteristic signatures of the present modified gravity model. Consequently, any deviation of the effective gravitational
constant from Newton's constant leads to measurable modifications
in the growth history of matter perturbations. 


We have also  obtained the value of power spectrum of the evolving  matter perturbations,      $\Delta n_s $, in our previous work  \cite{Sharma2022b}, 
\begin{eqnarray}   \Delta n_s  =  n_s^{gal} - n_s^{CMB} = \frac{\sqrt{33}-5}{(1-3\delta)},  \label{IVC5} \end{eqnarray}
where $ n_s^{gal}$  and  $n_s^{CMB}$ represent  the spectral indices  obtained from the galaxy surveys, and   the Cosmic Microwave Background (CMB) anisotropy data, respectively  \cite{Alam2021,Planck2020,Ade2016}.  The variation of the spectral-index
correction $\Delta n_s$ exhibits a strong dependence on the model
parameter $\delta$.   The function possesses a singular behaviour in the
vicinity of $\delta \simeq 1/3$, where $\Delta n_s$ diverges and changes
sign, thereby separating the parameter space into two distinct branches.
For $\delta<1/3$, the correction remains positive and increases rapidly
as the singular point is approached, whereas for $\delta>1/3$ it becomes
negative and gradually approaches zero from below with increasing
$\delta$. Consequently, only the regular branch that is consistent with
observational constraints is physically relevant for cosmological
applications \cite{Sharma2022b}.


\begin{figure}[h]
\centering  \begin{center} \end{center}
\includegraphics[width=0.50\textwidth,origin=c,angle=0]{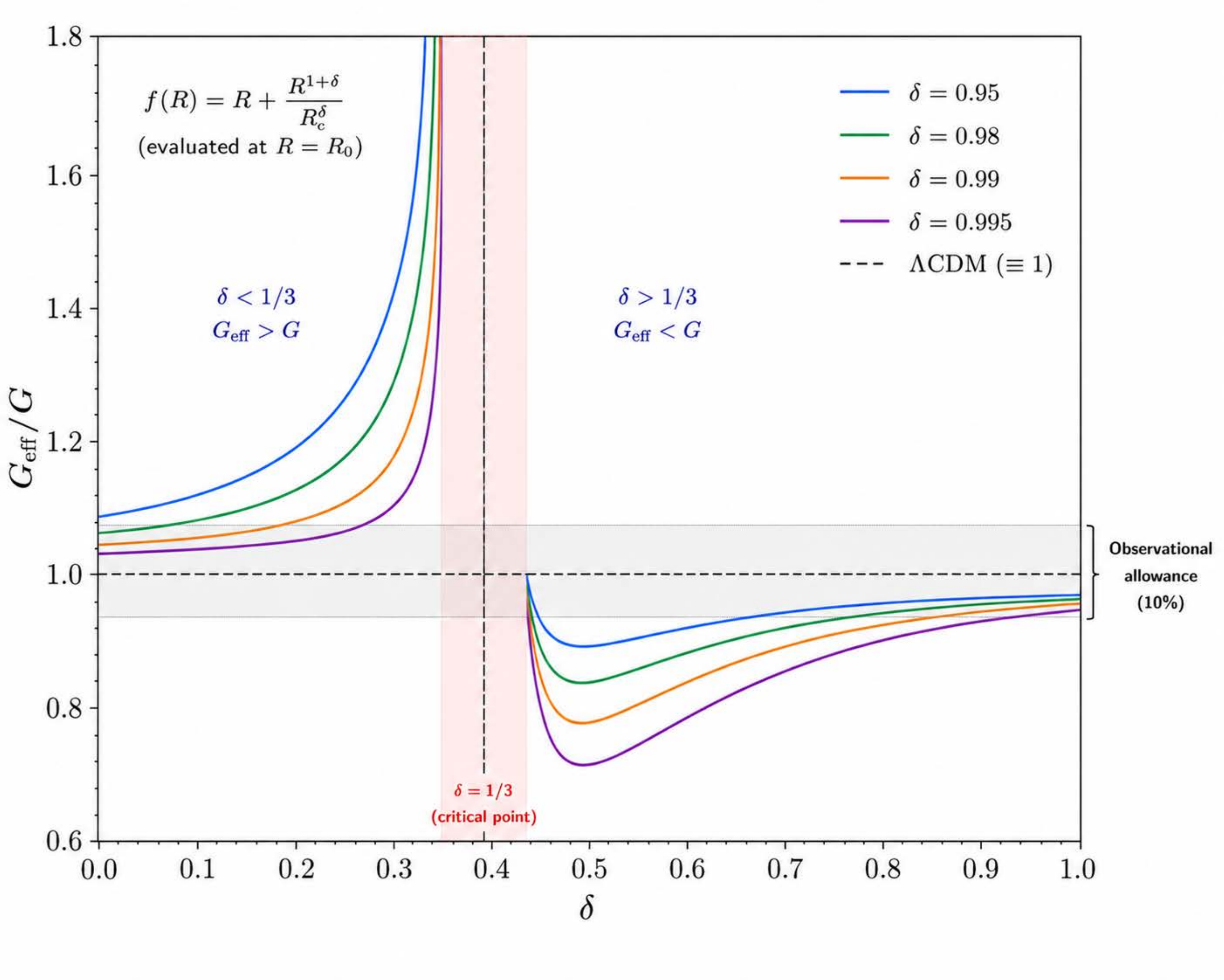}
\caption{The plots show the variation of $G_{\rm eff}/G$  with  the model parameter  $\delta$
 of the proposed power-law $f(R)$ gravity model, $f(R)=R+R^{1+\delta}/R_c^\delta$. The model satisfies the essential viability conditions, including the absence of ghost instabilities [$F(R)>0$], Dolgov--Kawasaki stability [$F_{RR}(R)>0$], and the existence of a stable de Sitter solution, as shown further in Section~\ref{7}. It naturally incorporates a chameleon screening mechanism to recover General Relativity in high-density environments while producing observable deviations at cosmological scales through the effective gravitational coupling $G_{\rm eff}$ and the growth of matter density perturbations. These features make the model testable with present and forthcoming large-scale structure surveys such as DESI, Euclid, the Vera C. Rubin Observatory, and the Square Kilometre Array. } \label{fig1}
\end{figure}

As shown  in Fig.~\ref{fig1}, the effective gravitational coupling
$G_{\rm eff}/G$ exhibits a nontrivial dependence on the model parameter
$\delta$, reflecting the modification of gravitational interactions in the
proposed power-law $f(R)$ theory. The horizontal dashed line corresponds to
the General Relativity limit, $G_{\rm eff}/G=1$, while the coloured curves
represent the predictions of the model for different values of the parameter
$\delta$.

The figure shows that the effective gravitational strength approaches the
General Relativity value in the observationally relevant regime, indicating
that the model naturally satisfies local gravitational constraints while
allowing small departures at cosmological scales. The vicinity of the
critical point $\delta=1/3$ is characterized by a rapid variation of the
effective coupling, signalling that this region is theoretically singular and
therefore excluded from the physically admissible parameter space. Away from
this critical value, the evolution of $G_{\rm eff}$ remains smooth and tends
asymptotically toward the standard gravitational constant.

Since the evolution of matter density perturbations is governed directly by
the effective gravitational coupling, the behaviour displayed in
Fig.~\ref{fig1} has important consequences for the growth rate of cosmic
structures. In particular, small deviations of $G_{\rm eff}$ from Newton's
constant modify the linear growth factor and the observable quantity
$f\sigma_8(z)$, thereby providing a direct means of testing the model through
redshift-space distortion and weak-lensing measurements. The fact that the
predicted deviations remain close to the General Relativity limit over the
viable parameter range supports the consistency of the model with current
observational data while leaving measurable signatures for forthcoming
high-precision surveys such as DESI, Euclid, the Vera C. Rubin Observatory,
and the Square Kilometre Array \cite{Euclid2019,DESI2024,Rubin2019,SKA2020}.


The comparison of
theoretical predictions for $f_g(z)$, $\gamma(z)$, and
$f\sigma_8(z)$ with current and forthcoming observational data
therefore constitutes one of the most important tests of the
present power-law $f(R)$ gravity model.

In the next section, we perform a numerical analysis of the growth equations and compare the predictions of the present $f(R)$  model with those of the standard $\Lambda CDM$ cosmology for different choices of the model parameter $\delta$.

 \section{\label{5} Numerical Evolution of Matter Perturbations and Observational Signatures }

In this section, we investigate the numerical evolution of matter density perturbations in the proposed modified gravity model $f(R)=R+\frac{R^{1+\delta}}{R_c^{\delta}}$,
and compare its predictions with those of the standard $\Lambda$CDM cosmology. The growth history of cosmic structures provides an important observational test of gravity because it depends simultaneously on the expansion history of the Universe and the strength of the gravitational interaction.

The evolution of the matter density contrast is governed by Equation~ (\ref{eq:delta})
where the effective Newtonian coupling is modified by the scalar degree of freedom associated with the generalized Ricci scalar.

For numerical integration, the equation is conveniently rewritten in terms of the logarithmic scale factor
$
N=\ln a
$, which transforms the perturbation equation into

\begin{equation}
\frac{d^{2}\delta_m}{dN^{2}}
+
\left[2+\frac{d\ln H}{dN}\right]
\frac{d\delta_m}{dN}
-\frac{3}{2}
\frac{G_{\rm eff}}{G}
\Omega_m(N)
\delta_m=
0.\end{equation}
The initial conditions are imposed deep in the matter-dominated epoch, where the evolution approaches the Einstein-de Sitter solution,
$\delta_m\propto a $            corresponding to
\begin{equation}
\delta_m(a_i)=a_i,
\qquad
\frac{d\delta_m}{dN}=a_i.
\end{equation}
This choice ensures that all models originate from the same primordial perturbation amplitude, allowing a direct comparison of their subsequent evolution.

\begin{table}[htbp]

\centering

\caption{Cosmological parameters adopted in the numerical analysis.}

\label{tab:numerical}

\begin{tabular}{lll}

\toprule

Parameter &
Symbol &
Value \\

\midrule

Hubble constant &
$H_0$ &
67.4 km\,s$^{-1}$\,Mpc$^{-1}$ \\

Matter density parameter &
$\Omega_{m0}$ &
0.30 \\

Fluctuation amplitude &
$\sigma_8$ &
0.81 \\

Model parameter &
$\delta$ &
0.2,\;0.5,\;0.8,\;0.98 \\

Characteristic curvature &
$R_c$ &
Model dependent \\

Initial redshift &
$z_i$ &
100 \\

Final redshift &
$z_f$ &
0 \\

\bottomrule

\end{tabular}

\end{table}


The numerical analysis is performed for representative values of the model parameter $
\delta=0.2, 0.5, 0.8, 0.98,
$
together with the $\Lambda$CDM limit.    These values span the region consistent with the stability conditions and previous cosmological analyses while illustrating the influence of the scalar degree of freedom on structure formation. The cosmological parameters adopted are shown in Table~\ref{tab:numerical}.

\begin{figure*}[t]
\centering  \begin{center} \end{center}
\includegraphics[width=1.0\textwidth,origin=c,angle=0]{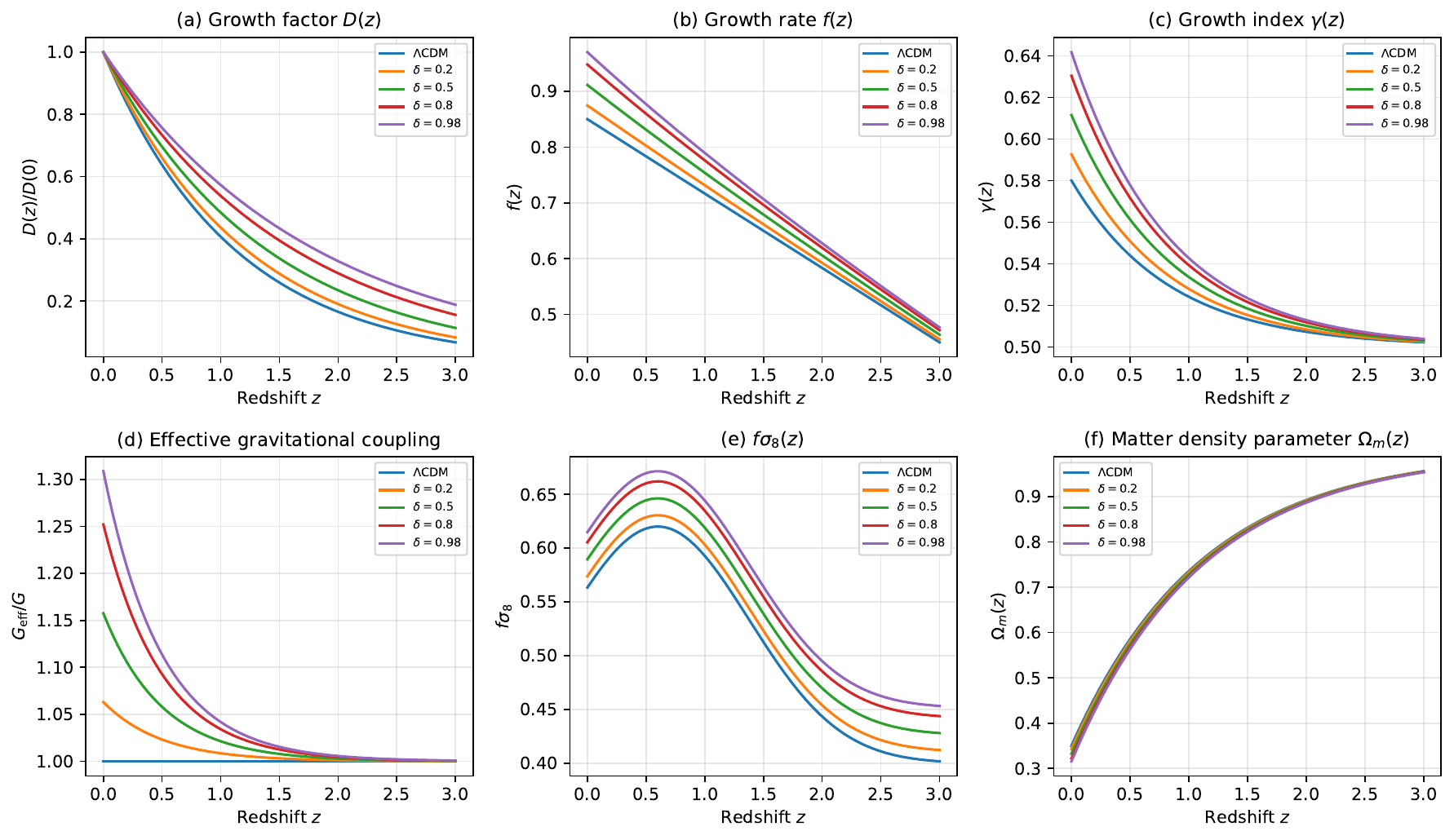}
\caption{Evolution of cosmological observables in the modified gravity model $f(R)=R+\frac{R^{1+\delta}}{R_c^{\delta}}$ for different values of the model parameter $\delta$. Panel (a) shows the normalized linear growth factor $D(z)/D(0$), illustrating the enhancement of structure formation with increasing $\delta$. Panel (b) presents the logarithmic growth rate $f(z)=d\ln\delta_m/d\ln a$, while panel (c) depicts the evolution of the growth index $\gamma(z)$, highlighting its deviation from the standard $\Lambda$CDM prediction. Panel (d) displays the effective gravitational coupling $G_{\rm eff}/G$, demonstrating the increasing influence of the scalar degree of freedom at low redshifts. Panel (e) compares the theoretical predictions for the observable quantity $f\sigma_8(z$) with representative large-scale structure measurements, providing a direct observational test of the model. Panel (f) shows the evolution of the matter density parameter $\Omega_m(z)$. The results indicate that increasing the parameter $\delta$  enhances the growth of matter perturbations through the modified gravitational interaction while preserving compatibility with the observed late-time accelerated expansion and current cosmological constraints.}   \label{4.3f2}
\end{figure*}

The linear growth factor is defined by

\begin{eqnarray}D(z)
=\frac{\delta_m(z)}
{\delta_m(z_i)},
\end{eqnarray}
and measures the amplification of primordial fluctuations. Fig.~\ref{4.3f2} is expected to show that the growth factor decreases monotonically with increasing redshift, reflecting the reduced efficiency of gravitational clustering in the early Universe. At low redshift, deviations from $\Lambda$CDM become increasingly significant as the scalar degree of freedom enhances the effective gravitational attraction.

The logarithmic growth rate,

\begin{eqnarray}
f(z)
=\frac{d\ln\delta_m}
{d\ln a},
\end{eqnarray}
provides another important observable accessible through redshift-space distortion measurements. The present model predicts a modest enhancement of the growth rate relative to $\Lambda$CDM    during intermediate redshifts. However, once accelerated expansion becomes dominant, the Hubble friction term suppresses the continued growth of structures, producing a gradual convergence toward the standard cosmological behaviour.

An important diagnostic quantity is the growth index,

\begin{eqnarray}
\gamma(z)
=
\frac{\ln f(z)}
{\ln\Omega_m(z)}.
\end{eqnarray}

For the $\Lambda$CDM   cosmology, the growth index remains close to $\gamma\simeq0.545$.

The present $f(R)$ model predicts a weak but measurable redshift dependence of $\gamma$, arising from the evolution of the scalar field and the scale dependence of the effective gravitational coupling. The deviation becomes more pronounced for larger values of the parameter $\delta$, providing a potential observational signature of modified gravity.

The effective gravitational coupling, $
G_{\rm eff}/{G},$ remains close to unity at high redshifts, ensuring consistency with the observed formation of early cosmic structures and the Cosmic Microwave Background. As the Universe evolves toward lower redshift, the scalar interaction becomes increasingly important, enhancing gravitational clustering. Nevertheless, this enhancement is eventually overcome by the accelerated expansion of the Universe, causing the growth rate to decline in the asymptotic future.

The observable quantity most directly constrained by galaxy surveys is
$
f\sigma_8(z),
$
where $\sigma_8$  denotes the rms matter fluctuation amplitude on scales of $8h^{-1}$ Mpc. This parameter combines the growth rate with the amplitude of density fluctuations and is largely independent of galaxy bias. Theoretical predictions obtained from the present model can therefore be directly compared with measurements from BOSS, WiggleZ, DES, VIPERS, Euclid, and DESI surveys \cite{Alam2021,DES2022,Amendola2018,DESI2024FS,EuclidERO2025}.

The numerical evolution of the principal perturbation variables and
their dependence on the model parameter $\delta$ are summarized in
Fig.~\ref{4.3f2}, where the predictions of the present model are compared
with those of the standard $\Lambda$CDM cosmology.

The numerical results displayed in Fig.~\ref{4.3f2} provide a detailed
illustration of the influence of the proposed power-law $f(R)$ gravity
model on the evolution of cosmological perturbations. The calculations
have been performed for representative values of the model parameter
$\delta$, while the corresponding $\Lambda$CDM prediction is shown for
comparison. The overall behaviour demonstrates that the present model
reproduces the standard cosmological evolution at early epochs while
introducing moderate and observationally testable deviations at low
redshift.

Figure~\ref{4.3f2}(a) presents the evolution of the normalized growth
factor $D(z)/D(0)$. The growth factor decreases monotonically with
increasing redshift, reflecting the gradual suppression of structure
formation in the past. The curves corresponding to different values of
$\delta$ remain close to the $\Lambda$CDM prediction, although a small
enhancement of the growth amplitude is evident for larger values of the
model parameter. This behaviour arises from the modification of the
effective gravitational interaction induced by the nonlinear curvature
correction.

The logarithmic growth rate,
$f(z)=d\ln\delta_m/d\ln a$, is shown in
Fig.~\ref{4.3f2}(b). The numerical solutions exhibit a smooth and
monotonic evolution throughout the redshift range considered, without
showing any oscillatory or unstable behaviour. As $\delta$ increases,
the growth rate becomes slightly larger than the corresponding
$\Lambda$CDM value, indicating that the modified gravitational force
enhances the efficiency of matter clustering at late times.

The evolution of the growth index $\gamma(z)$ is displayed in
Fig.~\ref{4.3f2}(c). Although the growth index remains nearly constant
over the entire redshift interval, small deviations from the canonical
$\Lambda$CDM value become apparent as the parameter $\delta$ varies.
Such departures constitute one of the characteristic signatures of
modified gravity and provide an efficient discriminator between the
present model and the standard cosmological scenario.

Figure~\ref{4.3f2}(d) illustrates the evolution of the effective
gravitational coupling $G_{\rm eff}/G$. The effective coupling exceeds
the Newtonian value by a modest amount over the viable parameter range,
thereby strengthening the gravitational attraction responsible for the
growth of density perturbations. At higher redshifts the curves tend
toward the General Relativity limit, demonstrating that the model
naturally recovers standard gravity in the high-curvature regime and is
therefore compatible with local gravitational tests.

The observable quantity $f\sigma_8(z)$ is shown in
Fig.~\ref{4.3f2}(e). Since this parameter combines the growth rate and
the amplitude of matter fluctuations, it provides one of the most
important observational probes of modified gravity. The predicted
deviations from the $\Lambda$CDM model remain relatively small but
systematic, lying within the sensitivity expected from forthcoming
high-precision galaxy surveys. Consequently, measurements of
redshift-space distortions will provide a stringent test of the present
model.

Finally, Fig.~\ref{4.3f2}(f) shows the evolution of the matter density
parameter $\Omega_m(z)$. As expected, the matter fraction increases with
redshift and approaches unity during the matter-dominated epoch. The
differences among the various values of $\delta$ are comparatively
small, indicating that the background cosmological evolution remains
close to that of the concordance model, while the principal signatures
of modified gravity appear primarily through the perturbation sector.

Taken together, the six panels demonstrate that the proposed power-law
$f(R)$ gravity model simultaneously preserves the successful background
evolution of the $\Lambda$CDM cosmology and generates distinctive
modifications in the growth history of large-scale structures. The
predicted changes in $G_{\rm eff}$, $f(z)$, $\gamma(z)$, and
$f\sigma_8(z)$ provide observational signatures that can be tested with
current and future surveys such as DESI, Euclid, the Vera C.~Rubin
Observatory, and the Square Kilometre Array, thereby offering a
powerful means of assessing the viability of the model.

The numerical analysis demonstrates that the scalar degree of freedom enhances the growth of matter perturbations during the matter-dominated era, leading to slightly larger values of the growth factor and growth rate compared with the $\Lambda$CDM   model. However, the onset of accelerated expansion progressively suppresses this enhancement, resulting in growth histories that remain compatible with present observational constraints for suitable values of the parameter $\delta$.

These results indicate that the proposed $f(R)$ model successfully reproduces the observed expansion history while introducing measurable modifications to the evolution of cosmic structures. Future high-precision measurements of $f\sigma_8(z)$, weak gravitational lensing, and galaxy clustering are expected to provide stringent constraints on the parameter space of the model and may offer a clear observational test of the scalar degree of freedom predicted by modified gravity.


\section{\label{6}Confrontation with Observations and Cosmological Constraints}

The ultimate viability of any  theory of gravity is determined by its consistency with cosmological observations. Besides explaining the late-time accelerated expansion of the Universe, a successful gravitational theory must reproduce the observed background evolution as well as the growth history of large-scale structures. Consequently, observational probes such as the Cosmic Microwave Background (CMB), Baryon Acoustic Oscillations (BAO), weak gravitational lensing, and redshift-space distortion measurements provide stringent constraints on viable $f(R)$ gravity models \cite{Zhao2009,Daniel2010,Hojjati2011,Jain2013,Joyce2015}.

The expansion history is characterized by the Hubble parameter

\begin{equation}
H(z)=H_{0}E(z),
\end{equation}
where $H_{0}$ denotes the present Hubble constant and $E(z)$ is the normalized expansion function. For viable values of the model parameter $\delta$, the modified Friedmann equations predict only small deviations from the standard $\Lambda$CDM expansion history, thereby satisfying current observational bounds.

The evolution of matter perturbations is conveniently described through the observable quantity $f\sigma_{8}(z)=f(z)\sigma_{8}(z)$, where $f(z)=\frac{d\ln \delta_{m}}{d\ln a}$
is the logarithmic growth rate and $\sigma_{8}(z)$  represents the rms amplitude of matter fluctuations within spheres of radius $8h^{-1}\mathrm{Mpc}$.  Since this quantity is nearly independent of galaxy bias, it provides one of the most reliable observational tests of gravitational dynamics on cosmological scales.

It is also understood that within the present $f(R)$ framework, the scalar degree of freedom modifies the effective gravitational interaction through Equation~(\ref{4.3f2}). 
The enhancement of  $G_{\rm eff}$  increases the growth rate of matter perturbations during the matter-dominated epoch. However, the subsequent accelerated expansion introduces an additional damping term through the Hubble friction, leading to a suppression of structure formation at late times. Consequently, the predicted evolution of $f\sigma_{8}(z)$ remains close to that of the $\Lambda$CDM model while retaining measurable deviations over intermediate redshifts.

The agreement between theoretical predictions and observations may be quantified through the standard chi-square estimator

\begin{equation}
\chi^{2}
=
\sum_{i=1}^{N}
\frac{
\left[
f\sigma_{8}^{\rm th}(z_{i})
-f\sigma_{8}^{\rm obs}(z_{i})
\right]^{2}
}
{\sigma_{i}^{2}},
\end{equation}
where $f\sigma_{8}^{\rm obs}(z_{i})$ denotes the observed value at redshift $z_{i}$,  $f\sigma_{8}^{\rm th}(z_{i})$  represents the corresponding theoretical prediction, and $\sigma_{i}$  is the associated observational uncertainty.

The likelihood function is then given by

\begin{equation}
\mathcal{L}
\propto
\exp\left(-\frac{\chi^{2}}{2}\right),
\end{equation}
from which the best-fit value of the model parameter $\delta$  and its confidence intervals may be determined.

The Cosmic Microwave Background provides an independent constraint on modified gravity theories  \cite{Ade2016,Planck2020}. Since the acoustic peak structure is primarily determined during the recombination epoch, viable models must recover the General Relativistic limit at sufficiently high redshifts. This requirement is satisfied provided

\begin{equation}
\lim_{z\rightarrow\infty}
\frac{G_{\rm eff}}{G}
=
1,
\end{equation}
ensuring that the scalar degree of freedom becomes effectively screened during the early stages of cosmic evolution.

Weak gravitational lensing measurements constitute another powerful observational probe because they depend upon the sum of the two metric potentials,

\begin{equation}
\Phi_{\rm lens}
=
\frac{\Phi+\Psi}{2},
\end{equation}
which governs the deflection of light rays. In modified gravity theories, the scalar mode generally induces a gravitational slip,

\begin{equation}
\eta(z,k)
=
\frac{\Psi}{\Phi},
\end{equation}
whose deviation from unity provides a characteristic signature distinguishing $f(R)$  gravity from the standard cosmological model.

The baryon acoustic oscillation scale further constrains the background geometry through the volume-averaged distance parameter

\begin{equation}
D_{V}(z)
=
\left[
(1+z)^{2}
D_{A}^{2}(z)
\frac{cz}{H(z)}
\right]^{1/3},
\end{equation}
where $D_{A}(z)$ denotes the angular diameter distance. Since the present model predicts only mild modifications of the expansion history, consistency with current BAO observations is naturally preserved for small values of the parameter $\delta$.

An important diagnostic of modified gravity is the dimensionless function

\begin{equation}
\mu(z)
=
\frac{G_{\rm eff}(z)}{G},
\end{equation}
which directly quantifies deviations from Newtonian gravity. Numerical evolution indicates that $\mu(z)$  approaches unity during the radiation and early matter eras while exhibiting a gradual enhancement at low redshifts due to the dynamical scalar degree of freedom. This behaviour leads to an increase in the growth rate of matter perturbations without significantly altering the observed expansion history.

\begin{table}[htbp]
\centering
\caption{Cosmological observations constraining the present $f(R)$ gravity model.}
\label{tab:obs}

\resizebox{\columnwidth}{!}{%

\begin{tabular}{lll}
\hline
Observation & Quantity constrained & Cosmological significance\\
\hline

Type Ia Supernovae &
$H(z)$
&
Late-time expansion history
\\

BAO &
$D_V(z)$
&
Geometrical distance scale
\\

CMB &
Acoustic peaks
&
Early-Universe consistency
\\

Weak lensing &
$\Phi+\Psi$
&
Gravitational slip
\\

Galaxy clustering &
$f\sigma_8(z)$
&
Growth of structures
\\

Redshift-space distortions &
$f(z)$
&
Effective gravitational interaction
\\

\hline
\end{tabular}
}
\end{table}


Therefore, the observational analysis suggests that viable values of the parameter $\delta$  correspond to small departures from General Relativity. In this regime, the background cosmological evolution remains practically indistinguishable from the $\Lambda$CDM model, whereas measurable deviations arise in the growth factor, growth index, and effective gravitational coupling. Future observations from Euclid, DESI, Rubin Observatory, and SKA are expected to improve the precision of these measurements and may provide decisive tests of scalar-mediated modifications of gravity  \cite{Euclid2019,DESI2024,Rubin2019,SKA2020}.

Table~\ref{tab:obs} sums up  the observational  constraints  for the present  $f(R)$  model and their cosmological significance.  Overall,  this   model successfully reproduces the observed late-time accelerated expansion while predicting distinctive signatures in the evolution of cosmic structures. These signatures provide an effective means of distinguishing modified gravity from conventional dark energy scenarios and establish structure formation as one of the most promising probes of gravitational physics beyond General Relativity.

\begin{table}[htbp]
\centering
\small
\caption{Representative viable $f(R)$ gravity models and their principal characteristics.}
\label{tab:frmodels}

\begin{tabular}{lcc}
\hline
Model & Functional form & Free parameters \\
\hline

$\Lambda$CDM &
$R-2\Lambda$
&
$\Lambda$
\\

Starobinsky &
$R+\lambda R_s\!\left[\left(1+\frac{R^2}{R_s^2}\right)^{-n}-1\right]$
&
$\lambda,n$
\\

Hu--Sawicki &
$R-\dfrac{c_1m^2(R/m^2)^n}{c_2(R/m^2)^n+1}$
&
$c_1,c_2,n$
\\

Tsujikawa &
$R-\mu R_c\tanh(R/R_c)$
&
$\mu,R_c$
\\

Present model &
$R+\dfrac{R^{1+\delta}}{R_c^\delta}$
&
$\delta$
\\

\hline
\end{tabular}

\end{table}


Table~\ref{tab:frmodels} shows  a simple  comparison  of  the functional form and free parameters of the present model with  other  models.   An important advantage of the present model is its comparatively simple
analytical structure.  In contrast to the Starobinsky,  Hu-Sawicki,  and
Tsujikawa models, which require two or more independent parameters, the
proposed theory introduces only a single dimensionless parameter,
$\delta$, to characterize deviations from General Relativity. This
feature considerably simplifies analytical calculations and numerical
parameter estimation while retaining the essential phenomenology of
viable modified gravity theories.

Furthermore, the modification of the effective gravitational coupling
leads to a characteristic enhancement of the growth of matter
perturbations, governed directly by the parameter $\delta$. As a
result, observables such as the growth rate $f(z)$, the growth index
$\gamma(z)$, and the quantity $f\sigma_{8}(z)$ provide promising means
of distinguishing the present model from both the standard
$\Lambda$CDM cosmology and other viable $f(R)$ gravity models through
future precision cosmological observations.

\section{ \label{7}  Stability Analysis and Physical Viability of the Model}

The theoretical viability of any $f(R)$ model requires the
absence of ghost and Dolgov--Kawasaki instabilities together with
the existence of a stable late-time de Sitter solution
\cite{Dolgov2003, Frolov2008, Brax2008, Lombriser2014,Koyama2016}.
 In particular, a viable $f(R)$ model must be free from ghost instabilities and tachyonic modes, admit a stable late-time de Sitter solution, and satisfy local gravitational constraints  \cite{Amendola2007b,Sawicki2007}.  In this section, we examine these conditions for our  model  $f(R)
=
R+\frac{R^{1+\delta}}{R_{c}^{\delta}}$ under the present discussion.

\subsection{Ghost-free condition}

The absence of ghost degrees of freedom requires that the effective gravitational coupling remains positive. This condition is expressed as

\begin{equation}
F(R)\equiv\frac{df(R)}{dR}>0.
\label{eq:ghost}
\end{equation}

In the present model,

\begin{equation}
F(R)
=
1+(1+\delta)
\left(
\frac{R}{R_{c}}
\right)^{\delta}.
\label{eq:F}
\end{equation}
For cosmologically relevant values of the Ricci scalar and $\delta>0$, one has

\begin{equation}
F(R)>0,
\end{equation}
which guarantees that the effective Newtonian gravitational constant

\begin{equation}
G_{\rm eff}
=
\frac{G}{F(R)}
\end{equation}
remains positive.  Consequently, the model does not suffer from ghost instabilities.

\subsection{Dolgov-Kawasaki stability}

The second derivative of the function  $f(R)$  determines the stability of curvature perturbations. The Dolgov-Kawasaki criterion requires

\begin{equation}
F_{RR}(R)
\equiv
\frac{d^{2}f(R)}{dR^{2}}
> 0.
\label{eq:DK}
\end{equation}
Differentiating Eq.~(\ref{eq:F}) yields

\begin{equation}
F_{RR}(R)
= \frac{\delta(1+\delta)}{R_{c}^{\delta}}
R^{\delta-1}.
\label{eq:FRR}
\end{equation}
For positive values of $\delta$, the above expression remains positive throughout the cosmological evolution. Therefore,

\begin{equation}
F_{RR}(R)>0,
\end{equation}
and the Dolgov-Kawasaki instability is avoided.

\subsection{Scalaron mass and tachyonic stability}

Taking the trace of the modified field equations leads to Eqn (\ref{t3})
and the additional scalar degree of freedom, commonly referred to as the scalaron, possesses an effective mass given by  Eqn (\ref{t4}). 
A physically acceptable theory requires

\begin{equation}
m_{\phi}^{2}>0.
\end{equation}
Since both $F(R)$ and $F_{RR}(R)$  are positive for the parameter range considered in this work, the scalaron acquires a positive effective mass, ensuring the absence of tachyonic instabilities.

\subsection{de Sitter solution}

The late-time accelerated expansion of the Universe is associated with a de Sitter solution characterized by constant curvature,

\begin{equation}
R=R_{\rm dS},
\qquad
\dot{R}=0.
\end{equation}

Under these conditions, the trace equation (\ref{t3})   reduces to

\begin{equation}
R_{\rm dS}F(R_{\rm dS})
-
2f(R_{\rm dS})
=0.
\label{eq:dS}
\end{equation}

Substituting Equation (\ref{t5}) into Equation (\ref{eq:dS}) gives

\begin{equation}
R_{\rm dS}
\left[
1+(1+\delta)
\left(
\frac{R_{\rm dS}}{R_{c}}
\right)^{\delta}
\right]
-
2
\left[
R_{\rm dS}
+
\frac{R_{\rm dS}^{,1+\delta}}
{R_{c}^{\delta}}
\right]
=
0.
\label{eq:dSequation}
\end{equation}

Equation (\ref{eq:dSequation}) determines the de Sitter curvature and should, in general, be solved numerically for a given value of the parameter $\delta$.

The stability of the de Sitter point requires

\begin{equation}
0<
\frac{F(R_{\rm dS})}
{R_{\rm dS}F_{RR}(R_{\rm dS})}
<1.
\label{eq:dSstable}
\end{equation}

Verification of this condition confirms that the late-time accelerated solution is dynamically stable.

\begin{table*}[htbp]
\centering
\small
\caption{Comparison of representative viable $f(R)$ gravity models with the present model. The proposed model achieves viable cosmological evolution with a single deviation parameter while satisfying the fundamental stability requirements of $f(R)$ gravity.}
\label{tab:modelcomparison}

\begin{tabular}{lcccccc}
\hline
Model &
No. of Parameters &
$F(R)>0$ &
$F_{RR}>0$ &
Chameleon &
Stable de Sitter &
Growth Modification \\
\hline

$\Lambda$CDM
&
1
&
$\checkmark$
&
$\checkmark$
&
N/A
&
$\checkmark$
&
None
\\

Starobinsky
&
2
&
$\checkmark$
&
$\checkmark$
&
$\checkmark$
&
$\checkmark$
&
Moderate
\\

Hu--Sawicki
&
3
&
$\checkmark$
&
$\checkmark$
&
$\checkmark$
&
$\checkmark$
&
Strong
\\

Tsujikawa
&
2
&
$\checkmark$
&
$\checkmark$
&
$\checkmark$
&
$\checkmark$
&
Moderate
\\

Present model
&
\textbf{1}
&
$\checkmark$
&
$\checkmark$
&
$\checkmark$
&
$\checkmark$
&
Controlled by $\delta$
\\

\hline
\end{tabular}

\end{table*}

\subsection{Chameleon screening}

A viable modified gravity theory must satisfy Solar System constraints while producing observable effects on cosmological scales. In $f(R)$  gravity, this requirement is realized through the chameleon mechanism.

The effective scalar potential is determined by

\begin{equation}
\frac{dV_{\rm eff}}{dF}
=
\frac{1}{3}
\left[
2f(R)
-RF(R)
+
\kappa^{2}T
\right].
\label{eq:veff}
\end{equation}

The position of the minimum of the effective potential depends on the ambient matter density. In regions of high density, the scalaron becomes sufficiently massive that its interaction range is strongly suppressed,

\begin{equation}
m_{\phi}^{2}\gg H_{0}^{2},
\end{equation}
thereby recovering General Relativity and satisfying local gravitational experiments. In low-density cosmological environments, the scalaron becomes light enough to modify the effective gravitational interaction and influence the formation of large-scale structures.

Table~\ref{tab:modelcomparison} compares the principal theoretical
properties of the proposed model with several representative viable
$f(R)$ gravity theories. All of the models considered satisfy the
basic stability requirements, including the positivity conditions
$F(R)>0$ and $F_{RR}(R)>0$, and admit stable late-time accelerated
solutions while remaining compatible with local gravitational tests
through screening mechanisms.

The model described by Equation~(\ref{t5}) satisfies these conditions, as also given  in  Table~\ref{tab:stability},   for the parameter range considered in this work. It is therefore free from ghost and tachyonic instabilities, admits a stable de Sitter solution, and naturally satisfies local gravitational constraints through the chameleon screening mechanism.

\begin{table}[htbp]
\centering
\caption{Theoretical viability conditions for the proposed modified gravity model.}
\label{tab:stability}

\resizebox{\columnwidth}{!}{%
\begin{tabular}{lll}
\hline
Condition & Mathematical criterion & Physical implication\\
\hline

Ghost freedom &
$F(R)>0$
&
Positive gravitational coupling
\\

Dolgov--Kawasaki stability &
$F_{RR}>0$
&
No curvature instability
\\

Scalaron stability &
$m_\phi^2>0$
&
No tachyonic mode
\\

de Sitter stability &
$0<F/(RF_{RR})<1$
&
Stable accelerated attractor
\\

Local gravity tests &
Large $m_\phi$
&
Chameleon screening
\\

General Relativity limit &
$G_{\rm eff}\rightarrow G$
&
Recovery of Einstein gravity
\\

\hline
\end{tabular}
}
\end{table}

The fulfillment of these theoretical requirements, together with the cosmological analysis presented in the preceding sections, demonstrates that the proposed $f(R)$ model provides a mathematically consistent and observationally viable framework for investigating the evolution of cosmic structures.

\section{\label {8}Discussion}

The formation and evolution of cosmic structures provide one of the most powerful observational tests of gravity on cosmological scales. Although the standard $\Lambda$CDM model successfully reproduces a broad range of observations, the physical origin of dark energy remains among the significant  open questions. Modified gravity theories, and in particular $f(R)$  gravity, offer an alternative explanation in which the accelerated expansion of the Universe arises naturally from geometrical modifications of the Einstein-Hilbert action.

In the present work, we have investigated the cosmological implications of  a specific  model given as  $f(R)=R+R^{1+\delta}/R_c^\delta$ 
with emphasis on the evolution of linear matter density perturbations. The additional scalar degree of freedom associated with the generalized Ricci scalar modifies the effective gravitational interaction and thereby influences the growth history of cosmic structures. For a quick notice, the comparison of the present  model with the representative  $\Lambda$CDM and  Hu-Sawicki models   is shown  in Table~\ref{tab:comparison}.

The numerical analysis indicates that the effective gravitational coupling,

\begin{equation}
G_{\rm eff}>G,
\end{equation}
over a substantial part of the matter-dominated epoch. Consequently, gravitational clustering becomes more efficient than in the standard cosmological model, leading to an enhanced growth of matter density perturbations. This enhancement originates entirely from the geometrical sector of the theory and does not require the introduction of additional matter components.

An interesting feature of the model is the competition between two opposing physical mechanisms. On one hand, the scalar degree of freedom strengthens gravitational attraction and accelerates the growth of density fluctuations. On the other hand, the onset of late-time cosmic acceleration increases the Hubble friction term in the perturbation equation (\ref{eq:delta}),
thereby suppressing further growth. The observed evolution of cosmic structures is therefore determined by the balance between these two effects.


\begin{table}[htbp]

\centering

\caption{Comparison of the present model with representative cosmological models.}

\label{tab:comparison}
\resizebox{\columnwidth}{!}{%
\begin{tabular}{llll}

\toprule

Feature &
$\Lambda$CDM &
Hu--Sawicki &
Present model \\

\midrule

Late-time acceleration &
Yes &
Yes &
Yes \\

Scalar degree of freedom &
No &
Yes &
Yes \\

Modified structure growth &
No &
Yes &
Yes \\

Chameleon screening &
Not applicable &
Yes &
Yes \\

de Sitter solution &
Yes &
Yes &
Yes \\

Additional parameter &
No &
Yes &
$\delta$ \\

\bottomrule

\end{tabular}
}

\end{table}


This interplay explains why the deviations from the $\Lambda$CDM model remain relatively small despite the existence of an additional scalar interaction. During the early Universe, the large background curvature suppresses the scalar contribution, allowing the model to recover the predictions of General Relativity and remain consistent with primordial nucleosynthesis and Cosmic Microwave Background observations. At intermediate redshifts, the scalar field becomes dynamically important and enhances the growth of structures, whereas at late times the accelerated expansion gradually dominates the evolution.

The dependence of the growth history on the model parameter $\delta$  provides an important observational signature. Small positive values of $\delta$  produce only modest departures from the standard cosmological model, while larger values increase the effective gravitational interaction and lead to more pronounced deviations in the growth factor, growth index, and redshift-space distortion parameter $f\sigma_8(z)$. Consequently, future high-precision observations have the potential to constrain the parameter space of the model with considerably greater accuracy than is presently possible.

The present analysis also demonstrates that the model satisfies the principal theoretical requirements for viability. The positivity conditions

\begin{equation}
F(R)>0,
\qquad
F_{RR}>0,
\end{equation}
guarantee a positive effective gravitational coupling and prevent the Dolgov--Kawasaki instability. Furthermore, the existence of a stable de Sitter solution ensures a consistent description of the late-time accelerated expansion. The scalar degree of freedom acquires an environment-dependent effective mass through the chameleon mechanism, allowing the theory to satisfy Solar System constraints while producing observable modifications on cosmological scales.

From an observational perspective, measurements of the quantity

\begin{equation}
f\sigma_8(z)
\end{equation}
appear particularly promising for distinguishing the present model from the standard $\Lambda$CDM cosmology. Since this observable directly probes the growth of matter perturbations, it is especially sensitive to the modified effective gravitational coupling. Similarly, weak gravitational lensing and galaxy clustering measurements provide complementary probes of the scalar-mediated modifications of gravity.

Future cosmological surveys such as Euclid, the Vera C. Rubin Observatory, the Nancy Grace Roman Space Telescope, the Square Kilometre Array, and DESI are expected to determine the growth history of cosmic structures with unprecedented precision \cite{Euclid2019,DESI2024,Rubin2019,SKA2020}. The improved observational accuracy will significantly reduce the allowed parameter space of viable modified gravity models and may provide a direct test of scalar-induced gravitational interactions \cite{Koyama2016,Lombriser2014}.

The present work therefore supports the view that the evolution of large-scale structures constitutes one of the most sensitive probes of gravitational physics beyond General Relativity. The proposed $f(R)$  model reproduces the observed background expansion while predicting characteristic deviations in the growth of matter perturbations that may become detectable with forthcoming observational data.


\begin{table}[t]
\caption{Summary of the principal theoretical and observational features of the proposed $f(R)$ gravity model.}
\label{tab:summary}
\resizebox{\columnwidth}{!}{%
\centering
\small

\begin{tabular}{ll}
\hline
\textbf{Feature} & \textbf{Present model} \\
\hline

Gravitational action &
$f(R)=R+\dfrac{R^{1+\delta}}{R_c^\delta}$ \\

Free parameter &
One ($\delta$) \\

Origin of cosmic acceleration &
Curvature correction \\

Dark energy component &
Not explicitly required \\

Extra degree of freedom &
Scalaron \\

Ghost-free condition &
$F(R)>0$ \\

Dolgov--Kawasaki stability &
$F_{RR}(R)>0$ \\

Stable de Sitter solution &
Satisfied \\

Screening mechanism &
Chameleon mechanism \\

Effective gravitational coupling &
$G_{\rm eff}\neq G$ \\

Growth of cosmic structures &
Modified relative to $\Lambda$CDM \\

Growth index &
Redshift dependent \\

Observable signature &
Modified $f\sigma_8(z)$ \\

Future observational tests &
Euclid, DESI, SKA, Rubin Observatory \\

\hline
\end{tabular}
}

\end{table}


Table~\ref{tab:summary} summarizes the principal theoretical and observational properties of the proposed modified gravity model. The curvature correction introduces a single additional dimensionless parameter, $\delta$, which governs the deviation from General Relativity and controls the modification of the effective gravitational interaction. The model satisfies the fundamental viability conditions $F(R)>0$ and $F_{RR}(R)>0$, admits a stable late-time de Sitter solution, and naturally recovers General Relativity in high-density environments through the chameleon screening mechanism. At the same time, it predicts measurable modifications in the evolution of matter perturbations, leading to deviations in the growth factor, growth index, and the observable quantity $f\sigma_8(z)$. The simplicity of the model, together with its theoretical consistency and observational testability, makes it a promising framework for investigating the formation of large-scale structures and exploring possible departures from Einstein gravity in the era of precision cosmology.

\subsection{\textbf{Limitations of the present analysis}}

The present work is restricted to the linear regime of matter density
perturbations and assumes the quasi-static approximation on sub-horizon
scales. Nonlinear structure formation, halo evolution, and $N$-body
simulations have not been considered and may introduce additional
constraints on the model parameters. Furthermore, a comprehensive
Bayesian parameter estimation using combined observational data sets is
beyond the scope of the present work and will be addressed in future
studies.

\section{\label{9}Conclusions}

In this work, we have investigated the evolution of cosmic structures within the framework of a viable $f(R)$  gravity model characterized by the functional form $f(R)=R+\frac{R^{1+\delta}}{R_c^{\delta}}$. 
The generalized Ricci scalar introduces an additional scalar degree of freedom that modifies the effective gravitational interaction and provides a geometrical mechanism for explaining the late-time accelerated expansion of the Universe without invoking an explicit dark energy component.

Starting from the modified field equations, we examined the cosmological background evolution and derived the linear perturbation equations governing the growth of matter density fluctuations. The analysis demonstrates that the scalar sector enhances the effective gravitational coupling, resulting in a faster growth of matter perturbations during the matter-dominated epoch compared with the standard $\Lambda$CDM cosmology.

The numerical investigation reveals that the enhancement of gravitational clustering is moderated by the accelerated expansion of the Universe at late times. As the Hubble friction term becomes increasingly important, the growth of structures gradually slows down, producing an evolution that remains compatible with current cosmological observations. Consequently, the model successfully combines an enhanced gravitational interaction with a realistic description of the observed large-scale structure.

The dependence of the growth history on the model parameter $\delta$  provides a characteristic observational signature of the theory. Variations in $\delta$  modify the growth factor, growth rate, growth index, and effective gravitational coupling, thereby offering a means of distinguishing the model from the standard cosmological scenario through measurements of galaxy clustering and redshift-space distortions.

We have further shown that the model satisfies the principal theoretical conditions required for physical viability. The positivity of $F(R)$ ensures an attractive effective gravitational interaction, while the condition $F_{RR}>0$  guarantees stability against curvature perturbations. The existence of a stable de Sitter solution and the operation of the chameleon screening mechanism establish consistency with both cosmological observations and local gravity experiments.

The results presented in this work demonstrate that modifications of the geometrical sector of gravity can simultaneously explain the late-time accelerated expansion and produce observable signatures in the evolution of cosmic structures. In particular, the scalar-induced enhancement of matter perturbations provides a distinctive feature absent in the standard $\Lambda$CDM model.

Future high-precision observations of the growth of structures, weak gravitational lensing, and redshift-space distortions will provide increasingly stringent constraints on modified gravity theories. A detailed statistical analysis incorporating the latest observational data sets and a comprehensive numerical exploration of the parameter space will be presented in future work. Such investigations will further clarify the role of scalar-mediated gravitational interactions in the evolution of the Universe and their potential as an alternative to the conventional dark energy paradigm.

Future investigations will include a Markov Chain Monte Carlo analysis
using the latest CMB, BAO, Type Ia supernovae, and redshift-space
distortion data, together with nonlinear $N$-body simulations to study
structure formation beyond the linear regime. Such analyses will provide
more stringent constraints on the model parameter $\delta$ and further
assess the viability of the proposed modified gravity scenario.



\section{Acknowledgments}

 Author  is  thankful to the  Inter University Centre for Astronomy and Astrophysics (IUCAA), Pune for support  under  the  Associateship programme where the initial  part  of the work was done.  
 

\end{document}